\documentstyle[12pt]{article}

\begin{document}

\title{Decay Anisotropy   of $e^+e^-$ Sources \\ from $pN$ and $pd$
collisions \footnote{Supported by BMFT and GSI Darmstadt} \vspace*{5mm} \\}
\author{E.L. Bratkovskaya\thanks{
Permanent address: Bogolubov Laboratory of Theoretical Physics,
Joint Institute for Nuclear Research, 141980 Dubna, Moscow Region, Russia},
W. Cassing, U. Mosel,\\
O.V. Teryaev$^*$, A.I. Titov$^*$ and V.D. Toneev$^*$ \vspace*{2mm} \\
\small \em Institut f\"ur Theoretische Physik, Universit\"at Giessen,
D-35392 Giessen, Germany \\
\small  \em $^*$ Bogolubov Laboratory of Theoretical Physics,
Joint Institute for Nuclear Research,\\
\small \em 141980 Dubna, Moscow region,  Russia }
\maketitle

\begin{abstract}
A full calculation of lepton-pair angular characteristics is carried
out for $e^+e^-$ pairs created in $pp$, $pn$ and $pd$ collisions at
intermediate energies. It is demonstrated that the proposed new
observable, the dilepton decay anisotropy, quite sensitively changes
for different sources and may be useful for their disentangling. The
relevance of the dilepton decay anisotropy is shown in the context of a
puzzling energy behavior for the ratio of the lepton yield from $pd$
to $pp$ reactions as observed at the BEVALAC. \\[2cm]
\noindent
Submitted to Phys. Lett. B.
\end{abstract}

\newpage
As is known since a long time  dileptons are quite attractive probes
since they provide almost direct information on hot and dense nuclear
matter during its evolution in heavy-ion collisions at BEVALAC/SIS and
SPS energies \cite{Mosel91}.  The information carried out by leptons
may tell us not only about the interaction dynamics of colliding
nuclei, but also on properties of hadrons in the nuclear environment or
on a possible phase transition of hadrons into a quark-gluon plasma.

However, there are a lot of hadronic sources for dileptons because the
electromagnetic field couples to all charges and magnetic moments. In
particular, in hadron-hadron collisions, the $e^+e^-$ pairs are created
due to the electromagnetic decay of time-like virtual photons which can
result from the bremsstrahlung process or from the decay of baryonic
and mesonic resonances including the direct conversion of vector mesons
into virtual photons in accordance with the vector dominance
hypothesis.  In the nuclear medium, the properties of these sources may
be modified and it is thus very desirable to have experimental
observables which would allow to disentangle the various channels of
dilepton production.

For a decomposition of $e^+e^-$ sources contributing to the dilepton
invariant mass spectra it seems to be natural to start from the study
of elementary nucleon-nucleon collisions and then to move successively
to nucleon-deuteron and more complicated systems. A step in this
direction has been done in Ref.~\cite{W93} where the $pd/pp$ dielectron
ratio was measured  in the $1-5$~GeV energy range. In contrast to naive
expectations, this $pd/pp$ ratio displays a puzzling beam-energy
dependence which is still a matter of debate.  There are several
interpretations of this effect: the {\em interferences} of different
channels \cite{Schafer}, the contribution of the {\em inelasticity}
effects \cite{Haglin94}, and of the {\em subthreshold $\eta$-meson}
production \cite{TKB95}.  Even by considering the new and more precise
data on the transverse momentum distributions of lepton
pairs~\cite{W93} the puzzle could not be resolved unambiguously.

Recently, we have proposed to use lepton pair angular distributions for
a distinction between different  sources \cite{BTT94,BSCMTT94}.
Indeed, the coupling of a virtual photon to hadrons induces a dynamical
spin alignment of both the resonances and the virtual photons.  One
thus can expect that the angular distribution of a lepton will be
anisotropic with respect to the direction of the dilepton (i.e. virtual
photon) emission. It has been shown that due to the spin alignment of
the virtual photon and the spins of colliding or decaying hadrons, this
{\em lepton decay anisotropy} turns out to be quite sensitive to the
specific production channel \cite{BTT94,BSCMTT94,Hag95} .

Following our previous work \cite{BTT94,BSCMTT94} we focus in this
letter on the study of the dilepton angular anisotropy as a way to
disentangle various $e^+e^-$ sources in $pN$ and $pd$ reactions.
Alongside with an extended computation of the Dalitz decay channel for
pseudoscalar mesons, we present the first full results for
nucleon-nucleon collisions at energies of a few GeV. By a simple
extension of these calculations to the nucleon-deuteron case we will
investigate if the anisotropy effect can help in solving the $pd/pp$
dilepton ratio puzzle.

As in  \cite{BTT94,BSCMTT94} we choose the polar $\theta$ and azimuthal
$\varphi$ angles of the momentum $\vec l_-$ of a created electron with
respect to the momentum $\vec q$ of a virtual photon to characterize
the decay anisotropy,  where $\vec l_-, \vec l_+$ are measured in the
rest frame of this virtual photon, i.e.  $\vec q \equiv \vec l_- + \vec
l_+ = 0$.  For comparing the shape of the angular distributions for
different channels, the differential cross section for dilepton
production in the channel $i$ may be presented in the form:

\begin{eqnarray}
S_i(M,\theta ) \equiv {d\sigma_i\over dM d\cos\theta}
= A_i (1 + B_i \cos^2\theta),
\label{in1}\end{eqnarray}
where $M$ is the invariant mass of a lepton pair $(M^2 = q_0^2-\vec q^2)$
and $A_i$ is defined by the normalization to unity of the angular
distribution of (\ref{in1}). The decay anisotropy coefficient for
channel $i$, i.e. $B_i$, then is given by
\begin{eqnarray}
B_i = {S_i(M,\theta=0^o) \over S_i(M,\theta=90^o)} - 1.
\label{in2}\end{eqnarray}
Since the coefficient $B_i$ is sensitive to the spin structure of
the interacting hadrons, it is in  general  a function of $M$ and the
masses of the hadrons  involved in the reaction.

The total differential cross section for proton--nucleon collisions
can be represented as a sum of differential cross sections for
all channels:
\begin{eqnarray}
{d\sigma^{pN} \over dM d \cos\theta} = \sum\limits_{i=channel}
{d\sigma_i\over dM d\cos\theta} =  A (1+B^{pN}\cos^2\theta).
\label{Ssum}\end{eqnarray}
For the total anisotropy coefficient we then get
\begin{eqnarray}
 B^{pN} = \sum\limits_{i=channel} <B_i^{pN}>,  \hspace*{1cm}
 <B_i^{pN}> = {\displaystyle  {\displaystyle d\sigma_i\over dM} \cdot
{\displaystyle B_i\over 1+{\displaystyle 1\over 3} B_i} \over
\displaystyle \sum\limits_i {\displaystyle d\sigma_i\over dM} \cdot
{\displaystyle 1\over 1+{\displaystyle 1\over 3} B_i}},
\label{B_isum}\end{eqnarray}
where the special weighting factors originate in the necessary
angle-integrations.
Thus, the anisotropy coefficient $B^{pN}$ is the sum of the
``weighted'' anisotropy coefficients ($<B_i>$) for each channel $i$
obtained by means of the convolution of $B_i$ with
the corresponding invariant mass distribution.

In a preceding study \cite{BSCMTT94} the anisotropy coefficients were
calculated  for the bremsstrah\-lung and $\Delta$--Dalitz decay channels
within a microscopic One-Boson-Exchange (OBE) model~\cite{Schafer}.
We will use the latter results in the present investigation without any
modification. Note that we take into account the interference terms
between bremsstrahlung and $\Delta$--channels for proton--nucleon and
proton--proton interactions.
As a result, the contributions of these two channels to (\ref{B_isum}) are
reduced to a single coherent term.

	For the Dalitz decay of an $\eta$-meson in Ref.~\cite{BTT94}
a first simple estimate $B_\eta= 1$ had been made appropriate for an
$\eta$-meson with vanishing momentum in the center-of-mass system (cms)
of the colliding hadrons.  In this case the direction of a virtual
photon $\vec q_{\eta}$ is not influenced by the $\eta$-meson momentum
$\vec{P}_{\eta}$ which simplifies the kinematical considerations. In
extension to \cite{BSCMTT94} we now include the full dynamics,
$\vec{P}_{\eta} \ne 0$, where the direction of the dilepton momentum in
the cms of the colliding nucleons, $\vec q$, does not coincide with
that in the eta rest frame $\vec q_{\eta}$; as a consequence the
anisotropy coefficient becomes a function of the invariant mass $M$ and
the collision energy $T_{lab}$.

To take this functional dependence into account the $\eta$--channel for the
lepton differential cross section (\ref{in1}) is written in the
following form:
\begin{eqnarray}
&& S_\eta (M,\theta)  = \int d\vec P_\eta \ {d\sigma^\eta\over d \vec P_\eta}
\ |T(M,\theta, \vec P_\eta)|^2,
\label{s1}\end{eqnarray}
where the averaged transition matrix element of the $\eta$--decay is given by
\begin{eqnarray}
|T(M,\theta, \vec P_\eta)|^2 \sim \int dV \ {1\over M^4} \
(\epsilon_{\mu\alpha\beta\gamma} q_\beta {q_1}_\gamma \
\epsilon_{\nu\alpha\rho\sigma} q_\rho {q_1}_\sigma) \ \cdot L_{\mu\nu},
\label{eta3}\end{eqnarray}
and $q, q_1$ are the four momenta of a virtual and real photon, respectively.
The integral is taken over the available phase-space volume.
The lepton tensor $L_{\mu\nu}$ is defined as
\begin{eqnarray}
L_{\mu\nu} = {\rm Tr} \ \hat l_- \gamma_\mu \hat l_+ \gamma_\nu,
\label{eta4}
\end{eqnarray}
where $\hat l = \gamma_\sigma l^\sigma$, $ \ l_-, l_+$ are the four
momenta of the leptons.

For the $\eta$ production cross section
${d\sigma^\eta / d \vec P_\eta}$
we use the phase-space oriented expression:
\begin{eqnarray}
{d\sigma^\eta\over d \vec P_\eta} \approx {1\over E_\eta} \
{\sqrt{M_x^2 - 4 m_N^2}\over M_x} \ T^2_\eta(P_\eta),
\label{seta}\end{eqnarray}
with $M_x^2 = (p_a + p_b - P_\eta)^2 = s - 2 \sqrt{s} E_\eta + m_\eta^2 $.
Here $ s=(p_a+p_b)^2$ is the center-of-mass energy squared and  $p_a,
p_b, P_\eta$ are the four-momenta of the colliding nucleons and the
$\eta$--meson, respectively, while  $E_\eta$ is the energy of the
$\eta$-meson in the cms of the colliding nucleons.  It is noteworthy
that the particular case considered in Ref.~\cite{BTT94} corresponds
formally to the substitution $d\sigma^\eta / d \vec P_\eta \sim \delta(
\vec P_\eta )$.  The modification of the phase-space formula
(\ref{seta}) is included in the term $T^2_\eta(P_\eta)$, which is the
production matrix element squared.  Here we use a simple
parametrization of the calculations presented in Ref.~\cite{Vetter91},
where the $\eta$-production cross section in nucleon--nucleon
collisions was calculated on the basis of an effective OBE model.

Results of our numerical calculations for $B_\eta$ at initial
energies from the $\eta$ production threshold to about 5~GeV are shown
in Fig.~\ref{fig1}. We find that the anisotropy coefficient
indeed depends on both the initial energy $E$ and the dilepton mass $M$.
$B_\eta$ approaches 1 only close to threshold or for $M \rightarrow  0$.
This behavior is quite natural because at the threshold energy we have
$|\vec P_\eta| \approx 0$; for low mass dileptons their velocity
is close to the velocity of light and thus the difference between the
two directions discussed above becomes negligible. If the invariant
mass of a virtual photon is close to its kinematical limit ($M\to m_\eta$),
the anisotropy coefficient approaches a lower limit of about $0.1\div 0.25$
for bombarding energies from 1.26 to 5 GeV.

In  Fig.~\ref{fig2} the dilepton invariant mass distributions
$d\sigma/dM$ from $pp$ and $pn$ interactions are shown at bombarding
energies of 1.26~GeV and 2.1~GeV, which are the necessary ingredients
for calculating the weighted anisotropy coefficients, Eq.~(\ref{Bpd}).
In Fig. \ref{fig2} the ``$\eta$'' denotes the contribution of the
$\eta$--channel the ``$\Delta$'' labels the contribution of the
$\Delta$--resonance term, while ``$Br$'' denotes the bremsstrahlung
channel. The dashed curves are the sum of bremsstrahlung and
$\Delta$--channel with interference.

Using the $pN$  cross sections for $\Delta$--production and
bremsstrahlung channels from Ref.~\cite{Schafer}, we are now in the
position to present the first estimate for the weighted coefficients
$<B_i(M)>$ for $pp$ and $pn$ collisions at the energies of 1.26 and
2.1~GeV (Fig.~\ref{fig3}). The notation is the same as in
Fig.~\ref{fig2}. The solid line denoted by ``$all$'' represents the sum
of bremsstrahlung, $\Delta$--channel with interference  and $\eta$--decay
source. At the energy of 1.26~GeV the anisotropy is determined by the
$\Delta$--decay since its contribution is dominating at this energy (cf.
Fig.~\ref{fig2}). The large contribution of the $\eta$--source at
2.1~GeV leads to the specific structure in $<B_i(M)>$ for $M\le
0.3$~GeV.

As mentioned above, a possible explanation for the experimental $pd/pp$
dilepton ratio is related to the contribution of leptons from
$\eta$--decay~\cite{TKB95}. Whereas the energy threshold of
$\eta$--production in  $pN$ interactions is about 1.26 GeV, in the
proton-deuteron case the effective threshold is lower due to the Fermi
motion of the constituent nucleons in the deuteron.  Thus at projectile
energies below $1.26$~GeV the contribution from the $\eta$--Dalitz decay
to  the dilepton spectrum may be significant for the $pd$ case, but
negligible for $pp$. Such an enhancement of the $\eta$-yield in $pd$
compared to $pp$ reactions has been also seen experimentally by the
PINOT collaboration \cite{Pinot}. In line with this argument we expect
larger values for the $pd/pp$ dilepton ratio at low energy, and
according to the calculations above one  can expect that subthreshold
$\eta$--production may manifest itself also in the angular
characteristics (anisotropy coefficients) of the created dileptons.

In a first order approximation the decay anisotropy coefficient
for $pd$ collisions may be written as the sum of $B_i$ for
$pp$ and $pn$ interactions. However, this approximation is not  sufficient
in the energy region close to the $\eta$--threshold.
To take the contribution to the dilepton spectrum from near-threshold
$\eta$--production into account we calculate the coefficient $B^{pd}$ as
\begin{eqnarray}
&& B^{pd} = \sum\limits_{k=p,n} \!\!\!\! <B_{Br,\Delta}^{kN}>
+ <B_\eta^{pd}>, \label{Bpd}\\
&& <B_\eta^{pd}> = {  \displaystyle{\displaystyle d\sigma_\eta^{pd}\over dM}
\cdot {\displaystyle B_\eta^{pN} \over \displaystyle 1+{\displaystyle 1
\over 3} B_\eta^{pN}}
 \over \displaystyle \sum\limits_{k=p,n}
{d\sigma_{Br,\Delta}^{kN}\over dM}\cdot
{\displaystyle 1\over 1 + {\displaystyle 1\over \displaystyle 3}
B_{Br,\Delta}^{kN}}
+ \ {d\sigma_\eta^{pd}\over dM} \cdot {\displaystyle 1\over 1
+ {\displaystyle 1\over \displaystyle 3} B_\eta^{pN}} }.
\label{Bpdeta}\end{eqnarray}
It should be stressed that in Eqs.~(\ref{Bpd}), (\ref{Bpdeta}) we take into
account the interference between bremsstrahlung and $\Delta$--channel for
$pp$ and $pn$ separately.

Fig.~\ref{fig4} shows the resulting weighted anisotropy coefficients
$<B_i(M)>$ for $pd$ collisions at 1.26 and 2.1~GeV.  The cross section
${d\sigma_\eta^{pd} / dM}$ is taken from Ref.~\cite{TKB95} and the
Paris deuteron wave function \cite{Paris}  is used for describing the
internal nucleon distribution in the deuteron.  As expected, at
threshold energy the total $<B_{all}(M)>$  coefficient for the $pd$
reaction differs substantially from that for $pp$ or $pn$ reactions due
to the near threshold $\eta$--decay source.  For higher energies (2.1~GeV)
 the shape of $<B_{all}(M)>$ is similar in all these cases. This
energy-dependent effect is more pronounced in Fig.~\ref{fig5} where the
ratio of the anisotropy coefficients for $pd$ to $pp$ reactions,
$R_B = B^{pd}/B^{pp}$, is represented at the energies of 1.26 and
2.1~GeV. The comparison of the solid (including the $\eta$--Dalitz
decay) and dashed curves (without the $\eta$--channel) illustrates the
relative importance of the $\eta$--decay contribution to the observable
decay anisotropy of the dileptons.

Thus, summarizing, the calculated anisotropy coefficients for $pp$,
$pn$ and $pd$ reactions support our suggestion in
Refs.~\cite{BTT94,BSCMTT94} that the dilepton decay anisotropy may
serve as an additional observable to discriminate the dilepton sources
by experimental means.

\vspace*{3mm}
This research is partly performed in the framework of the Grants
$N^{0}_{-}$ MP8300 and RFE300 from the International Science
Foundation and Russian Government. The work of V.D.T was supported by
Grant $N^o_-$ 3405 from INTAS (International Association for promotion
of cooperation with scientists from the independent states of the
former Soviet Union).

\newpage
\section*{Figure captions}

\begin{figure}[h]
\caption{The anisotropy coefficient $B_\eta$ for the $\eta$--channel
at initial energies from 1.26 to 4.9~GeV.}
\label{fig1}

\caption{ The cross sections $d\sigma/dM$ for $pp$ and $pn$ interactions
at 1.26 and 2.1~GeV bombarding energies.
The ``$\eta$'' denotes the contribution of the $\eta$--channel,
the ``$\Delta$'' labels the contribution of the $\Delta$--resonance term,
while ``$Br$'' denotes the bremsstrahlung channel. The dashed curves are
the sum of bremsstrahlung and $\Delta$--channel with interference.}
\label{fig2}

\caption{The weighted anisotropy coefficients $<B_i(M)>$
for $pp$ and $pn$ collisions at 1.26 and 2.1~GeV. The notation is
the same as in Fig.~\protect\ref{fig2}. The solid line denoted by
``$all$'' represents the sum of bremsstrahlung, $\Delta$--channel with
interference and $\eta$--decay contributions.  }
\label{fig3}

\caption{The weighted anisotropy coefficients
$<B_i(M)>$ for $pd$ interactions at 1.26 and 2.1~GeV.
The notation is the same as in Fig.~\protect\ref{fig3}. }
\label{fig4}

\caption{The ratio of anisotropy coefficients for $pd$ to $pp$ reactions
at energies of 1.26 and 2.1~GeV.
The solid lines  correspond to calculations taking
into account the $\eta$--decay contribution while the dashed lines
do not include the $\eta$--Dalitz decay.}
\label{fig5}
\end{figure}

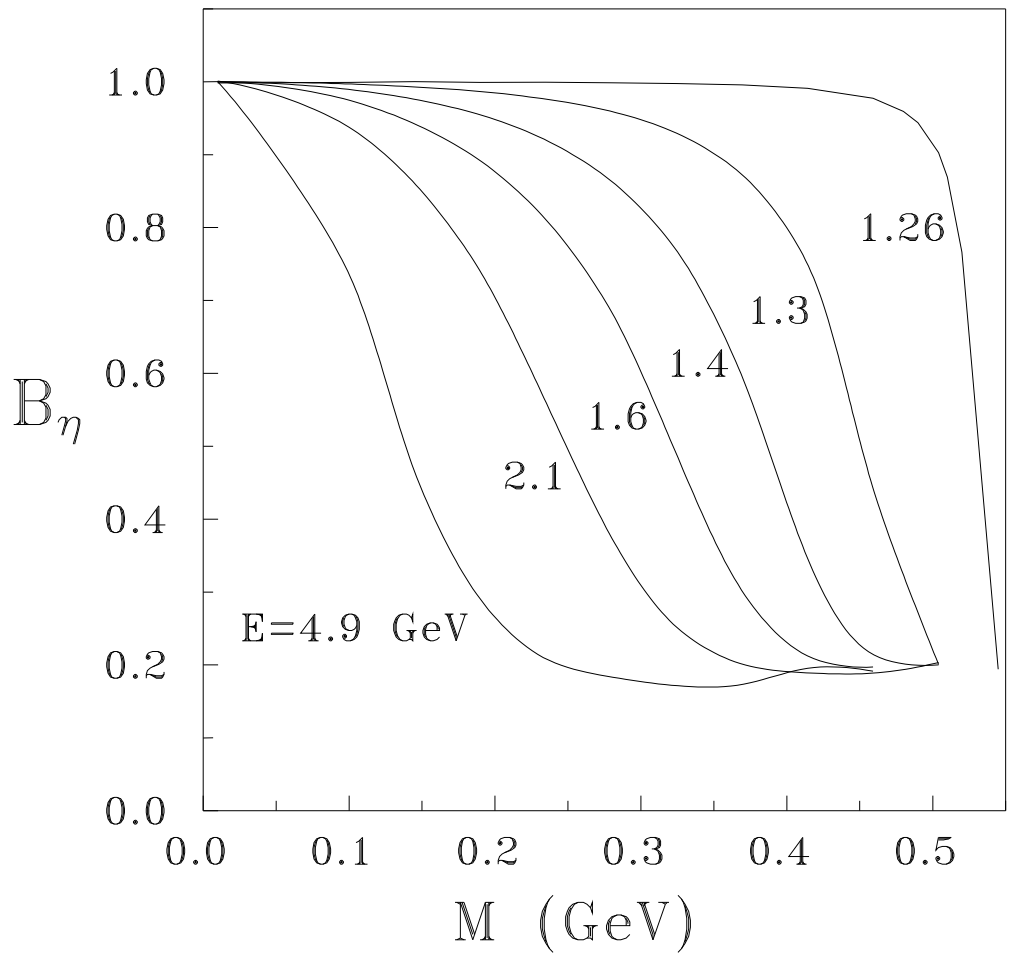
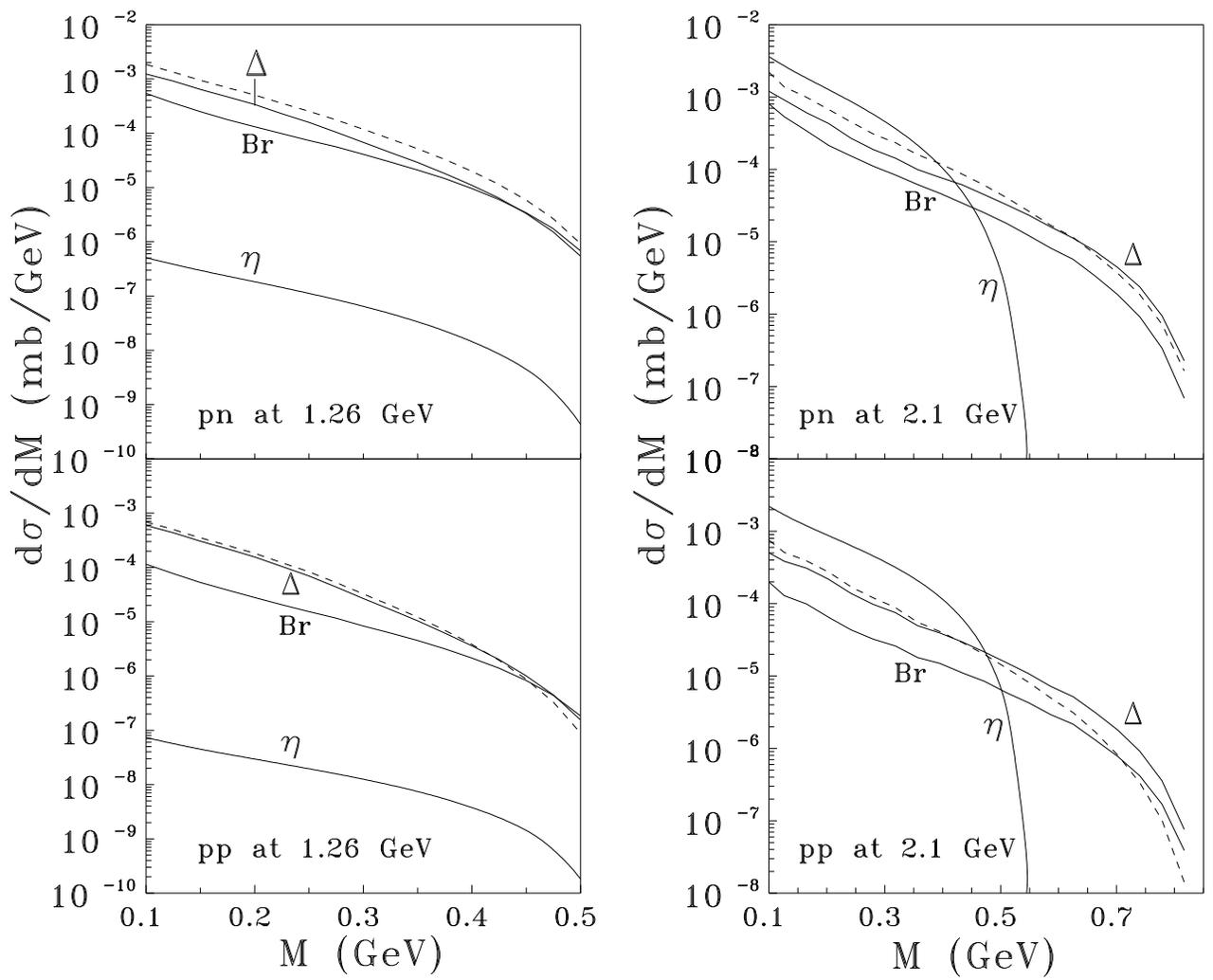
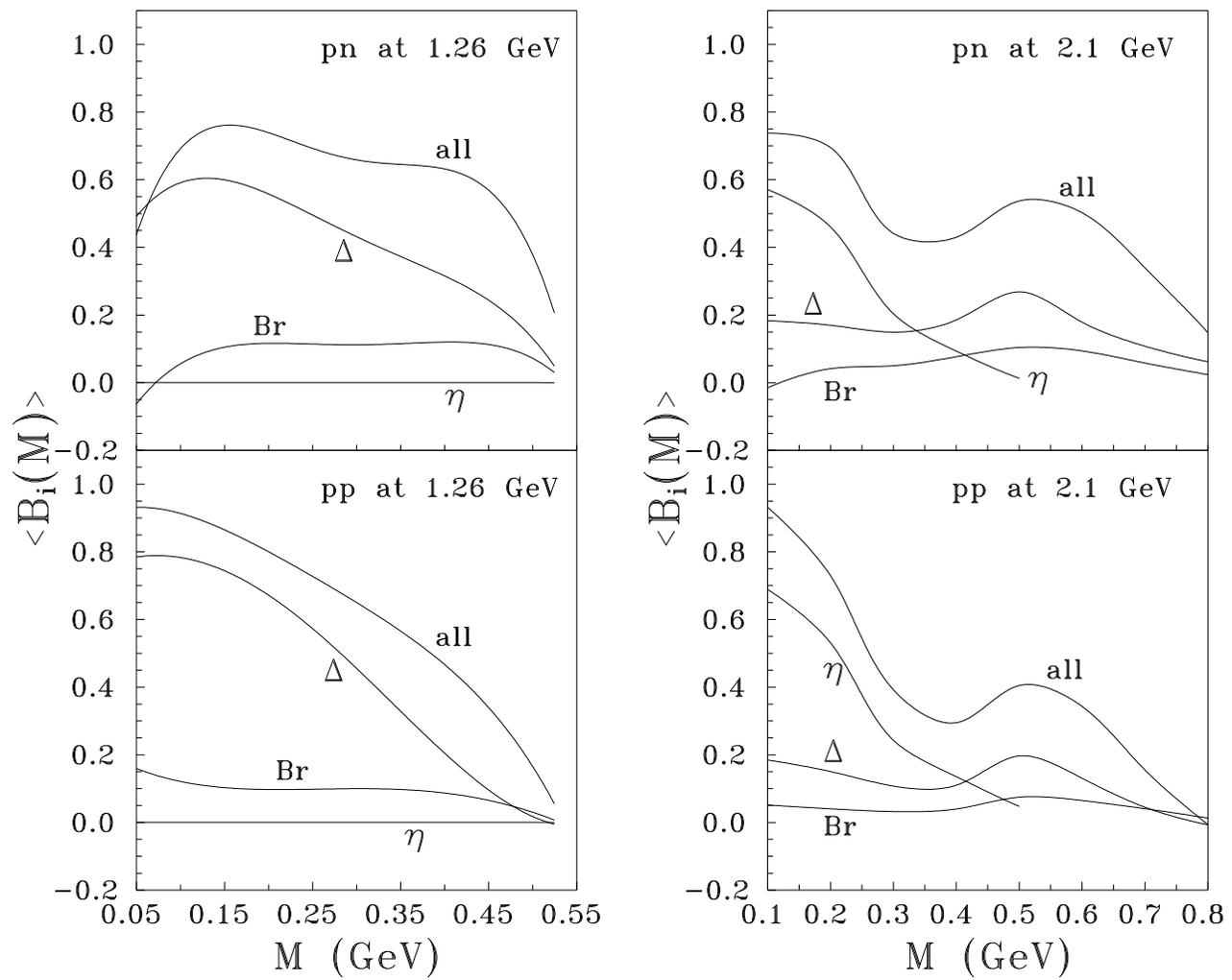
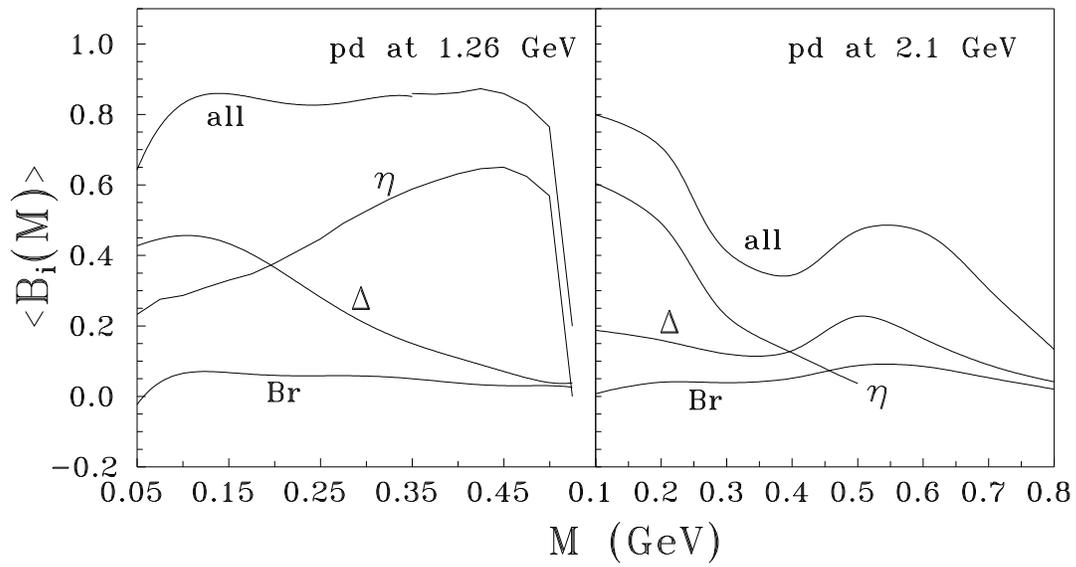
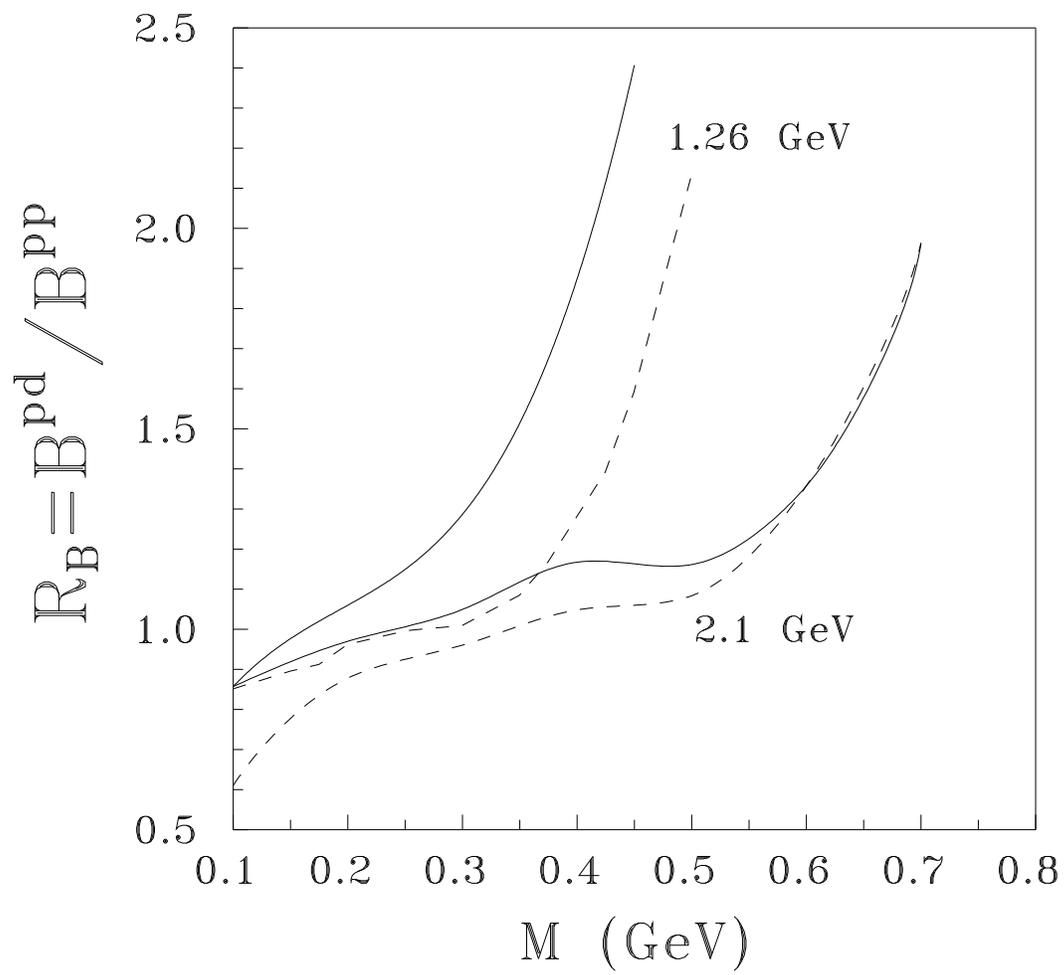
\end{document}